\documentclass[11pt]{article}

\usepackage[preprint]{acl}

\usepackage{times}
\usepackage{latexsym}

\usepackage[T1]{fontenc}

\usepackage[utf8]{inputenc}

\usepackage{microtype}

\usepackage{inconsolata}

\usepackage{graphicx}
\usepackage{placeins}
\usepackage{subcaption}
\usepackage{multirow}
\usepackage{booktabs}

\usepackage{algorithm}
\usepackage{algpseudocode}

\usepackage{amsmath}
\usepackage{amssymb}
\usepackage{mathtools}
\usepackage[table]{xcolor}
\usepackage[most]{tcolorbox}
\usepackage{xspace}

\usepackage[most]{tcolorbox}

\newtcolorbox{fedagentbox}[2][]{
  breakable,
  colback=gray!3,
  colframe=blue!45!black,
  colbacktitle=blue!45!black,
  coltitle=white,
  fonttitle=\bfseries,
  title={#2},
  boxrule=0.45pt,
  arc=1mm,
  left=1mm,
  right=1mm,
  top=1mm,
  bottom=1mm,
  #1
}

\title{XferAgent: Cross-Agent Experience Transfer for Self-Improving LLM Agents}

\author{
Weihao Li$^{1*}$, 
Jun Bai$^{2,3*}$, 
Shijie Chen$^{1}$,
Qincheng Lu$^{2,3}$, 
Xinyu Wang$^{2,3}$, 
Ziyang Song$^{4\dagger}$ \\[2mm]
$^{1}$McCormick School of Engineering, Northwestern University \\
$^{2}$School of Computer Science, McGill University \\
$^{3}$Mila -- Quebec AI Institute \\
$^{4}$School of Electrical Engineering and Computer Science, Ohio University \\[1mm]
}

\begin{document}
\maketitle

\begingroup
\renewcommand{\thefootnote}{}
\footnotetext{\hspace{0.0em}$^{*}$ Equal contribution.}
\footnotetext{\hspace{0.0em}$^{\dagger}$ Corresponding author: \texttt{ziyangs@ohio.edu}.}
\addtocounter{footnote}{-2}
\endgroup

\begin{abstract}
Large language model (LLM) agents can reuse experience from prior task executions to improve performance on subsequent tasks. Although existing methods enable cross-agent experience sharing, differences in agents' execution histories and task requirements make shared experiences difficult to reuse effectively. We propose \textbf{XferAgent}, a framework for \textbf{cross-agent experience transfer} that organizes reusable experience in local and shared experience graphs. XferAgent reconciles compatible contributions from heterogeneous agents while avoiding unreliable merges. For each new task, it selects shared experiences and integrates them with the target agent's local experience to support task execution. 
Experiments across diverse task domains show improvements over representative memory-based methods and effective experience transfer between heterogeneous agent frameworks. Cumulative task success also increases over successive task passes as agents update and reuse their experiences.

\end{abstract}

\section{Introduction}

Large language model (LLM) agents are increasingly used to solve complex tasks through multi-step reasoning, tool use, and interaction with external environments \citep{yao2022react, hong2024metagpt, qin2024toolllm}. Solving these tasks often requires iterative execution, exploration, and feedback-driven refinement before an agent reaches a successful outcome~\citep{shinn2023reflexion}. Across tasks, agents often encounter recurring reasoning patterns, tool-use procedures, and failure modes. Organizing these execution outcomes as reusable experience can help agents reuse effective strategies and avoid repeating previous failures on subsequent tasks. Recent work has therefore explored experience-driven self-improvement, where agents distill prior executions into reusable experience to improve performance on subsequent tasks~\citep{zhao2024expel, wang2025agent, cao2026remember}.

Existing memory-based approaches support local self-improvement by organizing experience from an agent’s past executions. Reflexion turns task feedback into self-reflection stored in episodic memory for subsequent attempts \citep{shinn2023reflexion}, while A-MEM dynamically links and updates memory notes as new experience arrives \citep{xu2026mem}. ExpGraph organizes reusable skills and lessons in an experience graph~\citep{feng2026expgraph}, while ReasoningBank distills successful and failed executions into reusable reasoning memories for subsequent tasks~\cite{ouyang2026reasoningbank}. These approaches establish that experience accumulated from past executions can improve subsequent behavior. However, they primarily support local experience reuse, where an agent can draw only on the strategies and failure lessons contained in its own execution history. Relevant strategies or failure lessons absent from one agent’s execution history may already have been acquired by other agents, motivating experience reuse beyond a single agent’s local history.

Recent work extends experience-driven self-improvement across agents by allowing one agent to reuse strategies and failure lessons learned by others. Agent KB aggregates trajectories from heterogeneous agent frameworks into a shared knowledge base \citep{tang2025agentkb}, while FoT aggregates reasoning experience from independently operating agents into a shared insight library~\citep{yao2026federation}. These studies show that agents can benefit from shared experience, but differences in reasoning protocols, tools, and execution histories make cross-agent experience transfer difficult~\citep{tang2025agentkb, xu2026rethinking, li2026fedagentke}. The central challenge is therefore to reconcile experience from heterogeneous agents into reusable shared experience and identify which shared experiences are useful for a new task while considering the target agent's local experience.

In this study, we propose \textbf{XferAgent}, a framework for \textbf{Cross-Agent Experience Transfer} across heterogeneous agents. XferAgent organizes each agent's reusable experience in a local experience graph and experience contributed across agents in a shared experience graph. Through \textit{Cross-Agent Experience Reconciliation}, XferAgent aligns compatible contributions while avoiding unreliable merges. For a new task, \textit{Task-Conditioned Experience Routing} selects a task-relevant subset of the Shared Experience Graph to construct a connected routed experience subgraph. 
\textit{Local-Shared Experience Integration} combines the routed subgraph with the target agent's local experience to select task-relevant guidance. Following task evaluation, \textit{Experience Updates} support subsequent experience reuse. Our main contributions are:

\begin{itemize}
    \item We formulate cross-agent experience transfer in experience-driven self-improving LLM agents and introduce XferAgent to support experience reuse across heterogeneous agents.

    
    \item We develop Cross-Agent Experience Reconciliation to align compatible contributions from heterogeneous agents while avoiding unreliable merges.

    
    \item We combine Task-Conditioned Experience Routing and Local-Shared Experience Integration to select task-relevant experiences using both local and cross-agent outcome evidence.

    \item Experiments across diverse tasks and LLM backbones demonstrate improvements over representative memory-based methods, with additional evidence of cross-framework experience reuse and increasing cumulative task success across repeated executions.

\end{itemize}

\section{Related Work}


\paragraph{Experience-Driven Agent Learning and Structured Memory.}
Experience-driven self-improvement enables LLM agents to leverage knowledge from prior executions without updating model parameters. Reflexion~\citep{shinn2023reflexion} uses verbal feedback to guide subsequent attempts, ExpeL~\citep{zhao2024expel} distills reusable insights from execution trajectories, and ReasoningBank~\citep{ouyang2026reasoningbank} extracts strategies from successful and failed experiences. More structured approaches organize such experiences through evolving memory representations. A-MEM~\citep{xu2026mem} dynamically links related memories, AriGraph~\citep{anokhin2025arigraph} integrates semantic and episodic knowledge into a graph-based world model, and ExpGraph~\citep{feng2026expgraph} organizes skills and lessons through graph-based retrieval and utility-aware ranking. However, these methods primarily focus on experience reuse within individual agents.


\paragraph{Cross-Agent Experience Sharing and Transfer.}
Recent studies explore cross-agent experience reuse through shared knowledge representations. Agent KB~\citep{tang2025agentkb} aggregates experiences from heterogeneous agent frameworks into a shared knowledge base, while FICAL~\citep{wu2024fical} exchanges distilled knowledge for federated in-context learning. FoT~\citep{yao2026federation} further distills distributed reasoning experiences into reusable metacognitive insights. These approaches facilitate cross-agent experience sharing, but experiences from heterogeneous agents may be difficult to align, and not all shared experiences are useful for a new task. XferAgent addresses these challenges by reconciling contributed experiences and integrating task-relevant shared experiences with the target agent's local experience.

\section{XferAgent}\label{sec:method}

\begin{figure*}[t]
\centering
\includegraphics[width=\textwidth]{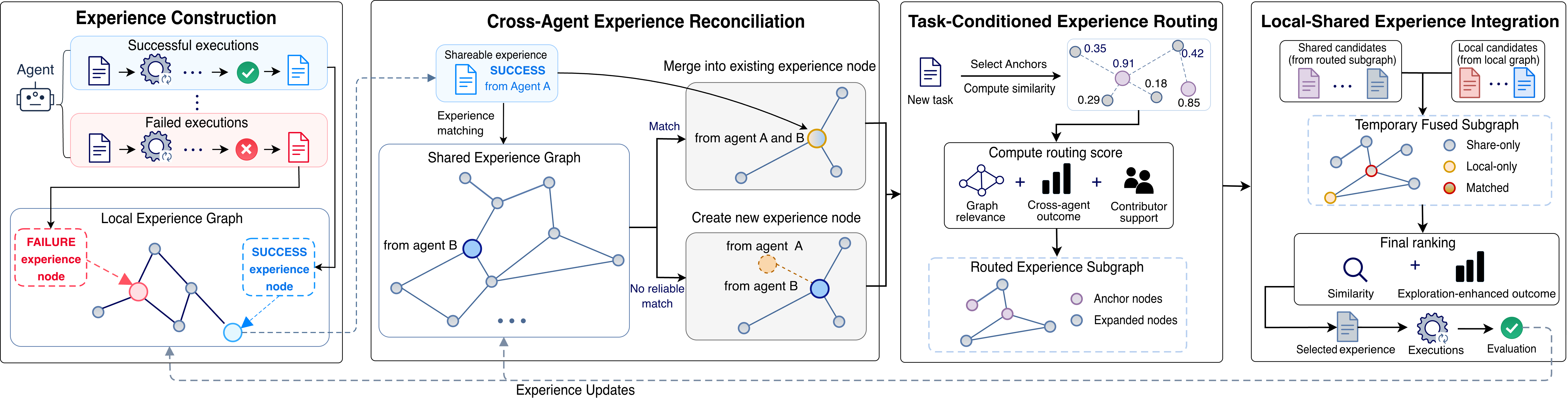}
\vspace{-1\baselineskip} 
\caption{\textbf{Overview of XferAgent.}
Agents construct local experience graphs by extracting experiences from successful and failed task executions. Cross-Agent Experience Reconciliation incorporates contributed experiences into a shared experience graph. Task-Conditioned Experience Routing selects task-relevant anchors and constructs a  routed experience subgraph. Local–Shared Experience Integration combines local and shared candidates into a temporary fused graph and ranks experiences for task execution, followed by evaluation-based experience updates.
}
\vspace{-1\baselineskip} 
\label{fig: overview}
\end{figure*}

\subsection{Problem Setup}

We consider $N$ heterogeneous agents, indexed by $i\in\{1,\ldots,N\}$. Agent $i$ maintains a local experience graph $\mathcal{G}_{i}^{L,t}$, while experience contributed across agents is organized in a shared experience graph $\mathcal{G}^{S,t}$, where $t$ denotes the task step.
Each node $v$ contains experience text $e_v$, a type $k_v\in\{\textsc{success},\textsc{failure}\}$, an embedding $h_v=\phi(e_v)$ computed by a text encoder $\phi$, provenance, and outcome statistics. 
Edges represent semantic similarity and typed relations between experiences. 

For a new task $x_t$ assigned to agent $i$ at time $t$, XferAgent retrieves experience from the graph state at $t-1$, before the task is executed. After task evaluation produces an outcome score $r_t$, XferAgent updates outcome statistics only for experiences included in the agent's prompt for that task. The completed execution is also converted into new reusable experience for subsequent tasks.

As shown in Figure~\ref{fig: overview}, XferAgent enables cross-agent experience transfer through Cross-Agent Experience Reconciliation (Section~\ref{sec:experience_reconciliation}), Task-Conditioned Experience Routing (Section~\ref{sec:Task-Conditioned Experience Routing}), Local-Shared Experience Integration (Section~\ref{sec:local-shared expereince integration}), and Experience Updates (Section~\ref{sec:experience_updates}).

\subsection{Cross-Agent Experience Reconciliation}
\label{sec:experience_reconciliation}

XferAgent converts each completed agent execution into reusable experience using a schema-constrained mapper. A successful execution produces a \textsc{success} experience, while an unsuccessful execution produces a \textsc{failure} experience describing a failed strategy or required check rather than the incorrect answer itself. The local experience node may retain richer execution details, such as tool observations and intermediate planning details. When the experience is contributed to the shared experience graph, XferAgent removes task-specific and execution-sensitive information, including the raw task, final answer, and complete trajectory. For a contributed experience, a stable identifier links the local experience node to its representation in the shared experience graph.

To organize reusable experiences into a graph, XferAgent connects each experience to a sparse set of semantically related experiences. For each node, it retains its \(K_{\mathrm{edge}}\) most similar experiences whose cosine similarity exceeds \(\theta_{\mathrm{edge}}\). A semantic edge is retained if either node selects the other, with cosine similarity used as its weight.


XferAgent reconciles each contributed experience before incorporating it into the shared experience graph. This is necessary because experience contributions from heterogeneous agents are not naturally aligned and may contain overlapping or semantically similar content. If a stable identifier matches an existing node in the shared experience graph, XferAgent merges the contributed experience with that node directly.
Otherwise, XferAgent searches the shared experience graph for potential matches with the same experience type and sufficient semantic similarity. For a contributed experience $v$, the candidate set is
\begin{equation}
\mathcal C(v)=\{u\in\mathcal V^{S}: k_u=k_v, \cos(h_u,h_v)\geq\theta_{\mathrm{rec}}\},
\label{eq:alignment_candidates}
\end{equation}
where $\theta_{\mathrm{rec}}$ is the minimum cosine similarity for a candidate match. The reconciliation rule considers three cases: (1) If no candidate exists, XferAgent creates a new node. (2) If exactly one candidate exists, XferAgent merges the contributed experience with that node. (3) If multiple candidates exist, XferAgent merges the contributed experience with the most similar candidate only when its cosine similarity exceeds that of the second-most similar candidate by at least a similarity margin \(\delta_{\mathrm{rec}}\); otherwise, it creates a new node to avoid an unreliable merge. When experiences are merged, their provenance is combined, while outcome statistics remain separate by agent. After reconciliation, XferAgent constructs semantic edges among the resulting nodes in the shared experience graph.

Beyond the undirected semantic edges, XferAgent also preserves typed relations between experiences. XferAgent adds a \textsc{DERIVED-FROM} relation from each newly contributed experience to every shared experience included in that task's prompt. These relations have a fixed weight of 1.


\subsection{Task-Conditioned Experience Routing}
\label{sec:Task-Conditioned Experience Routing}

For a new task $x_t$, XferAgent constructs a task-relevant routed experience subgraph instead of delivering the entire shared experience graph. Agent $i$ receives the identifiers and embeddings of available shared experience nodes and computes their task similarity locally as 
$a(x_t,v)=\cos(\phi(x_t),h_v)$. It selects up to a fixed number of the highest-scoring positive matches above $\theta_{\mathrm{match}}$ as anchors, which serve as starting points for graph search. Only the selected identifiers and similarity scores are sent for subgraph construction, while the raw task and its embedding remain with the agent.

Direct cosine similarity may miss experiences that are connected to relevant anchors but have low similarity to the task description. XferAgent therefore propagates task relevance from the anchors over the shared experience graph. The positive anchor scores are normalized to sum to one, defining $\boldsymbol{\rho}_i(x_t)$, while other nodes receive zero weight. We use personalized PageRank (PPR) to compute graph relevance scores \citep{Page1999ThePC}:
\begin{equation}
\mathbf s_i^{\mathrm{graph},(\ell+1)}
=\alpha\boldsymbol{\rho}_i(x_t)
+(1-\alpha)A^\top
\mathbf s_i^{\mathrm{graph},(\ell)},
\label{eq:view_pagerank}
\end{equation}
where $A$ is the row-normalized weighted adjacency matrix of the shared experience graph, $\alpha$ is the restart probability that balances anchor relevance and propagation through graph edges, and $\ell$ indexes the iterations. After iterations, $s^{\mathrm{graph}}_{i,v}$ denotes the graph relevance score of experience $v$, corresponding to its component in the resulting vector $\mathbf{s}_i^{\mathrm{graph}}$.

XferAgent also considers previous task outcomes when selecting shared experiences, since graph relevance alone does not capture past task performance. For target agent $i$, XferAgent computes a cross-agent outcome score from other agents' past task outcomes. Each experience $v$ in the shared experience graph stores a usage count $n^S_{jv}$ and a cumulative task outcome score $z^S_{jv}$ for each agent $j$. These statistics cover only tasks in which experience $v$ was included in agent $j$'s prompt.
The leave-one-agent-out (LOAO) calculation excludes agent $i$'s own outcomes. Therefore, XferAgent aggregates these statistics over all  other agents $j\neq i$:
\begin{align}
& n^{S}_{-i,v}=\sum_{j\neq i}n^{S}_{jv}, \quad
z^{S}_{-i,v}=\sum_{j\neq i}z^{S}_{jv}, \nonumber \\
& s^{\mathrm{outcome}}_{-i,v} =
\frac{z^{S}_{-i,v}+\kappa u_0}
     {n^{S}_{-i,v}+\kappa},
\label{eq:loao_utility}
\end{align}
where $u_0$ is the prior outcome score and $\kappa$ controls the strength of this prior.

XferAgent also considers contributor support $s^{\mathrm{support}}_{-i,v}$, which counts the number of distinct agents other than $i$ that contributed experience $v$. Repeated contributions from the same agent count only once. To favor less-used experiences, XferAgent adds an exploration bonus to the cross-agent outcome score based on other agents' usage counts. The resulting exploration-enhanced cross-agent outcome score is capped at one and denoted by $s^{\text{exp-outcome}}_{-i,v}$. XferAgent combines the graph relevance score, the exploration-enhanced outcome score, and the contributor support score to compute the final routing score:
\begin{align}
s^{\mathrm{route}}_{i,v}
={}&\beta_g\operatorname{Norm}
(s^{\mathrm{graph}}_{i,v})
+\beta_o\operatorname{Norm}
(s^{\text{exp-outcome}}_{-i,v})
\nonumber \\
&+\beta_s\operatorname{Norm}
(s^{\mathrm{support}}_{-i,v}),
\label{eq:routing_prize}
\end{align}
where $\operatorname{Norm}$ applies min-max normalization separately to each term over the available shared experience nodes. The fixed weights $\beta_g$, $\beta_o$, and $\beta_s$ control graph relevance, exploration-enhanced outcome, and contributor support, respectively.
Computational details appear in Appendix~\ref{app:experience_algorithm}. 

XferAgent uses the routing scores to construct a connected routed experience subgraph. It starts from the anchor with the highest task similarity and repeatedly selects the highest-scoring unselected neighbor of the selected nodes. Each candidate is scored using its routing score plus a fixed-weight bonus for its strongest edge to a selected node. Expansion stops when the node budget $B$ is reached or no eligible neighboring nodes remain. The resulting subgraph contains the selected shared experiences, their connecting edges, embeddings, and cross-agent outcome scores. If no qualifying anchor exists, XferAgent returns an empty subgraph.


\subsection{Local-Shared Experience Integration}
\label{sec:local-shared expereince integration}

XferAgent integrates the routed experience subgraph with the target agent's local experience graph to select experiences for the task prompt.
XferAgent first selects candidate experiences separately from the local experience graph and the routed experience subgraph, using task-matching nodes as anchors. It applies PPR to each graph using its own edges and retains a bounded number of nodes with the highest positive graph relevance scores. If the routed experience subgraph is empty, the agent can still select experiences from its local experience graph.

XferAgent then builds a temporary fused graph to combine local and shared candidates.  For each local candidate, it first checks for a match using its recorded shared identifier.  If no identifier match exists, XferAgent merges it with the highest-similarity experience of the same type already in the pool, provided their cosine similarity meets the fusion threshold. The pool includes previously processed local experiences. If no qualifying match exists, the experience remains separate. When a local experience matches a shared experience, it retains the richer local content while retaining the shared identifier for cross-agent outcome statistics.

XferAgent retains the graph edges among selected candidates and adds semantic links between unmatched local and shared experiences that meet the similarity threshold. It then applies PPR to the fused graph and retains a bounded number of candidates for final ranking. The fused graph is used only for the current task and does not modify either persistent experience graph. Merging matched experiences and adding semantic links can change the structure of the temporary fused graph. XferAgent therefore applies PPR again and retains a bounded number of nodes with the highest positive graph relevance scores for final ranking. The fused graph does not modify the structure of either original graph. Appendix~\ref{app:experience_algorithm}  provides details on edge construction, seed selection, and tie-breaking.

When the target agent has limited outcome history for an experience, outcomes from other agents provide an initial estimate. As local observations accumulate, they receive greater weight. Before executing task $x_t$, agent $i$ reads how often each local experience was included in its prompt, $n_{iv}^{L,t-1}$, and the sum of the corresponding task outcome scores, $z_{iv}^{L,t-1}$. For a candidate matched across both graphs, agent $i$ uses the cross-agent outcome score $s_{-i,v}^{\mathrm{outcome},t-1}$  as a shared prior and combines it with these local statistics:
\begin{equation}
s^{\mathrm{fused}}_{i,v} =
\frac{z_{iv}^{L,t-1}+\kappa s^{\mathrm{outcome}}_{-i,v}}
     {n_{iv}^{L,t-1}+\kappa},
\label{eq:local_global_utility}
\end{equation}

Here, $s^{\mathrm{fused}}_{i,v}$ denotes the fused outcome score for $x_t$, and $\kappa$ controls the weight of the shared prior relative to local observations. A local-only candidate uses the default prior $u_0$ while retaining its local usage count and cumulative outcome score. A shared-only candidate uses the cross-agent outcome score directly. All statistics are taken from the pre-task state $t-1$.

XferAgent ranks the remaining candidates using direct task similarity and the fused outcome score. To favor less-used experiences, it adds the exploration bonus from Section~\ref{sec:Task-Conditioned Experience Routing} to the fused outcome score. The resulting exploration-enhanced fused outcome score is capped at one and denoted by $s^{\text{exp-fused},t}_{i,v}$. For matched candidates, the exploration count combines local usage with other agents' shared usage. Otherwise, only the available count is used. The final selection score is defined as follows:
\begin{align}
    s^{\mathrm{select}}_{i,v} = 
    &(1-\lambda_u)\operatorname{Norm}(a(x_t,v)) \nonumber \\
    &+\lambda_u\operatorname{Norm}(s^{\text{exp-fused}}_{i,v}).
\label{eq:experience_ranking}
\end{align}
Here, $a(x_t,v)$ denotes task similarity. When candidates are merged, XferAgent retains the highest similarity score without recomputing it for the selected local content. The two terms are independently min-max normalized over the candidates selected by PPR on the fused graph. If a term has zero range, its values are clipped to the unit interval without rescaling (see  Appendix~\ref{app:experience_algorithm}). The fixed weight $\lambda_u$ controls their balance.  XferAgent selects experiences in descending score order, subject to limits on the number of experiences and their total text length. XferAgent selects no experiences when no usable seed exists in the fused graph or the highest score is below a fixed threshold. Selected experiences are included as \textsc{success} or \textsc{failure}.

\subsection{Experience Updates}
\label{sec:experience_updates}

Before executing task $x_t$, XferAgent records the local and shared experiences included in agent $i$'s prompt. After task evaluation produces an outcome score $r_t$, XferAgent updates each recorded experience's usage count as $n_v^t=n_v^{t-1}+1$ and cumulative outcome score as $z_v^t=z_v^{t-1}+r_t$. XferAgent updates only experiences included in the prompt, modifying agent $i$'s local statistics and its corresponding records in the shared experience graph. XferAgent then converts the completed execution into reusable experience as described in Section~\ref{sec:experience_reconciliation}. The richer experience is added to the local experience graph, while its shareable representation is reconciled before entering the shared experience graph.  XferAgent also links the new shared experience to each shared experience included in the task prompt through a \textsc{DERIVED-FROM} relation. These relations have a fixed weight of 1.

\FloatBarrier

\section{Experiments}

\subsection{Experimental Setup}
\label{sec:experiment_design}

\paragraph{Datasets and Backbones.}
We evaluate XferAgent primarily on GAIA~\citep{mialon2024gaia} and ALFWorld~\citep{shridhar2020alfworld}, which assess complementary agent capabilities: multi-step reasoning and tool use versus interactive task execution. GAIA contains 301 test tasks across three difficulty levels, evaluated with Qwen3-Next-80B-A3B-Instruct and GPT-5.6 Luna. ALFWorld comprises 274 tasks (140 seen and 134 unseen), evaluated with Qwen3.8-27B. 
To examine experience reuse in coding, we include a supplementary evaluation on MBPP~\citep{austin2021program} (60 coding tasks), using Llama-3.1-8B-Instruct.

\paragraph{Agent and Memory Configurations.}
For GAIA, we use SmolAgents~\citep{smolagents} for the main evaluation and additionally consider OWL~\citep{hu2025owl} for cross-framework experience transfer. For ALFWorld, agents interact with the environment through its text-based action interface. MBPP employs three logical agents with fixed task assignments and a shared LLM backbone. Across ALFWorld and MBPP, we use a remote-memory budget of $B=8$, at most 10 final experience items, and a nominal 2,400-token memory allowance. 
Memory construction and retrieval settings for these two benchmarks are detailed in Appendix~\ref{app:supplementary_protocols}.

\paragraph{Baselines.}
We compare XferAgent with No Memory and five representative memory-based methods. No Memory evaluates the same base agent without experience retrieval, providing a reference for assessing the benefits of memory augmentation. The memory-based baselines include AgentKB~\citep{tang2025agentkb} for cross-agent experience sharing, ReasoningBank~\citep{ouyang2026reasoningbank} for reusable reasoning memories, ExpGraph~\citep{feng2026expgraph} for graph-structured experience retrieval, Mem0~\citep{chhikara2025mem0} for persistent memory management, and LightMem~\citep{fang2026lightmem} for efficient memory consolidation. For cross-agent evaluation, we use benchmark-specific AgentKB configurations and adapt the other memory systems to support bounded retrieval across agent-owned memories. Implementation details for the supplementary-benchmark adapters are provided in Appendix~\ref{app:supplementary_protocols}.

\paragraph{Evaluation Protocols and Metrics.}
For the main benchmark comparisons, we evaluate the reuse of pre-constructed experience memories, keeping memory content and historical outcome statistics fixed during testing. Within each benchmark, methods use the same reasoning backbone and evaluation cohort, with benchmark-specific memory construction and retrieval settings. We separately examine Experience Updates over three task passes on 165 GAIA validation tasks, reporting cumulative task recovery as memories are updated after task evaluation. Performance is measured by accuracy on GAIA, success rate on ALFWorld, and program pass rate on MBPP. Failed or unparseable cases are included in the evaluation denominators. Further details on GAIA Test evaluation and the other two benchmark protocols are provided in Appendices~\ref{app:gaia_test_backbones} and~\ref{app:supplementary_protocols}, respectively.


\subsection{Main Results}
\label{sec:main_results}



The main experiments assess cross-agent experience transfer across diverse tasks and LLM backbones, covering question answering (GAIA), interactive task execution (ALFWorld), and code generation (MBPP). Table~\ref{tab:main_overall} presents the overall performance of XferAgent against representative memory-based methods.



\paragraph{GAIA Results.}
On GAIA (301 test tasks), XferAgent achieves the highest overall accuracy across both backbones. With GPT-5.6 Luna, it reaches 52.16\%, exceeding No Memory by 3.99 percentage points and the strongest memory-based baseline, ExpGraph, by 4.65 points. With Qwen3-Next-80B-A3B-Instruct, XferAgent achieves 29.57\%, compared with 28.90\% for AgentKB and 27.57\% for No Memory. Notably, all competing memory-based methods underperform No Memory with Luna, indicating that experience reuse does not necessarily improve task performance. Difficulty-level results are provided in Appendix~\ref{app:gaia_test_backbones}.




\begin{table*}[t]
\centering
\caption{Overall performance across GAIA, ALFWorld, and MBPP (\%). Metrics are accuracy, task success rate, and program pass rate, respectively. ALFWorld and MBPP follow benchmark-specific single-run protocols (Appendix~\ref{app:supplementary_protocols}). 
Bold indicates the best result in each column.}
\label{tab:main_overall}
\setlength{\tabcolsep}{10pt}
\renewcommand{\arraystretch}{0.9}
\begin{tabular*}{\textwidth}{@{\extracolsep{\fill}}lcccc@{}}
\toprule
& \multicolumn{2}{c}{\textbf{GAIA}}
& \textbf{ALFWorld}
& \textbf{MBPP} \\
\cmidrule(lr){2-3}
\cmidrule(lr){4-4}
\cmidrule(lr){5-5}
\textbf{Method}
& \shortstack{Qwen3-Next-80B\\A3B-Instruct}
& GPT-5.6 Luna
& Qwen3.8-27B
& \shortstack{Llama-3.1-8B\\Instruct} \\
\midrule
No Memory     & 27.57 & 48.17 & 84.7 & 50.0 \\
AgentKB       & 28.90 & 46.51 & 88.3 & 50.0 \\
ReasoningBank & 26.91 & 46.18 & 85.0 & 46.7 \\
ExpGraph      & 27.24 & 47.51 & 88.3 & 48.3 \\
Mem0          & 25.91 & 44.85 & 84.7 & 45.0 \\
LightMem      & 21.59 & 41.86 & 81.8 & 40.0 \\
\midrule
\textbf{XferAgent}
& \textbf{29.57}
& \textbf{52.16}
& \textbf{88.7}
& \textbf{53.3} \\
\bottomrule
\end{tabular*}
\end{table*}



\paragraph{ALFWorld Task Completion.}
On 274 ALFWorld tasks, XferAgent achieves the highest success rate of 88.7\%, outperforming No Memory (84.7\%) by 4.0 percentage points and slightly exceeding AgentKB and ExpGraph (both 88.3\%). This extends cross-agent experience transfer to multi-step interactive task execution. Implementation details, including AgentKB's retrieval of up to eight cross-agent candidates, are provided in Appendix~\ref{app:supplementary_protocols}.


\paragraph{MBPP Code Generation.}
On 60 MBPP tasks, XferAgent achieves the highest pass rate of 53.3\%, compared with 50.0\% for both No Memory and AgentKB, and 48.3\% for ExpGraph. Although the performance margin is relatively modest, the results provide complementary evidence of effective experience reuse in code generation. Together with GAIA and ALFWorld, these comparisons demonstrate the applicability of XferAgent across different task settings, with varying improvements over existing memory-based approaches.

\subsection{Progressive Self-Improvement through Experience Evolution}
\label{sec:cross_round}


To examine progressive self-improvement through experience evolution, we conduct three task passes over 165 GAIA validation tasks. After each execution, agents update their experience memories based on evaluation feedback for subsequent reuse. An independently evaluated No Memory baseline serves as a reference.

\begin{figure}[t]
\centering
\captionsetup{skip=5pt}
\includegraphics[width=0.98\linewidth]{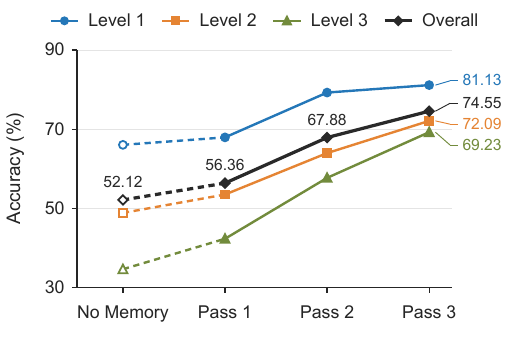}
\caption{
Cumulative task performance through experience evolution on GAIA validation (165 tasks). Passes 1--3 report cumulative task success across repeated executions. The independent No Memory baseline is connected to Pass 1 by dashed segments for visual comparison. Overall accuracy is computed over all 165 tasks.
}
\label{fig:round_evolution}
\end{figure}


Across three task passes, cumulative overall accuracy increases from 56.36\% to 67.88\% and 74.55\%, compared with 52.12\% for No Memory (Figure~\ref{fig:round_evolution}). Gains occur across all difficulty levels, with Level 3 improving from 42.31\% to 69.23\%. These results reflect cumulative success on repeated tasks rather than generalization to unseen tasks, and the independent No Memory baseline does not control for repeated attempts.



\subsection{Cross-Framework Experience Transfer}
\label{sec:framework_scaling}


We examine cross-framework experience transfer between SmolAgents and OWL with 2, 4, and 8 agents on GAIA validation. The initial two agents use different frameworks, while additional agents are randomly assigned to either framework as the group expands. Each agent builds experiences from 20 questions sampled from a shared pool of 105, yielding 40, 80, and 160 agent--question learning instances with possible overlap. All configurations are evaluated on the same 60 held-out questions (18, 29, and 13 at Levels 1--3).

\begin{figure*}[t]
\centering
\includegraphics[width=\textwidth]{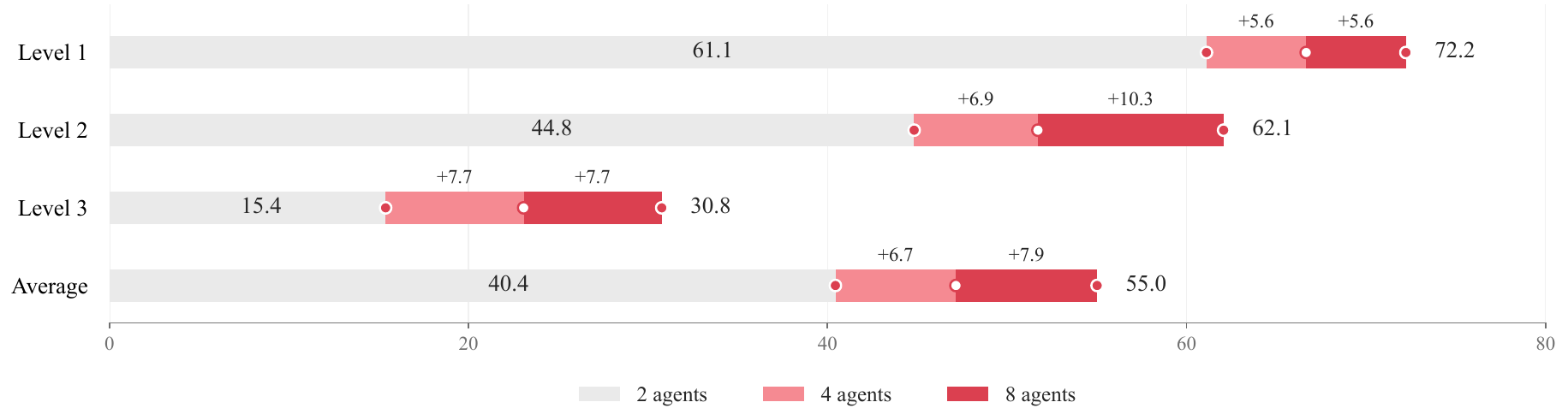}

\caption{Cross-framework experience transfer with 2, 4, and 8 agents on 60 held-out GAIA validation questions. Gray bars represent the 2-agent baseline, while light and dark red segments indicate accuracy gains from 2 to 4 and 4 to 8 agents, respectively. Average denotes the unweighted mean of accuracy across the three difficulty levels.}

\label{fig:cross_framework_agent_scales}
\end{figure*}


The macro-average accuracy increases from 40.4\% with 2 agents to 47.2\% with 4 agents and 55.0\% with 8 agents (Figure~\ref{fig:cross_framework_agent_scales}). Gains occur across all difficulty levels, with Level~3 improving from 15.4\% to 30.8\%. Since agent count and total learning exposure increase together, their individual effects cannot be isolated, nor can the effect of framework heterogeneity. Sampling details are provided in Appendix~\ref{app:cross_framework_protocol}.

\subsection{Ablation Studies}
\label{sec:ablation}

We conduct ablation studies on XferAgent using GAIA with GPT-5.6 Luna to assess the contributions of different experience sources and selection mechanisms. As shown in Table~\ref{tab:ablation_layout}, \textit{Local Experience Only} disables retrieval from the Shared Experience Graph, reducing overall accuracy from 52.16\% to 46.84\%, while \textit{Shared Experience Only} excludes local experience retrieval and achieves 50.83\%. 
This difference indicates that local and shared experiences provide complementary information, with local experience contributing more strongly in this setting.
Furthermore, \textit{w/o Outcome-Based Ranking} removes outcome-based scoring from experience ranking, decreasing accuracy to 48.17\%. \textit{w/o Connectivity Constraint} relaxes the connectivity requirement during shared experience selection, yielding 46.84\%. These findings highlight the contributions of outcome-informed selection and connected subgraph routing to effective experience reuse.

\begin{table}[t]
\centering
\caption{Ablation study of XferAgent on GAIA (301 test tasks) using GPT-5.6 Luna. Overall accuracy is reported in percentages, and $\Delta$ denotes the change relative to the full system in percentage points.}
\label{tab:ablation_layout}
\small
\setlength{\tabcolsep}{3pt}
\renewcommand{\arraystretch}{1.08}
\begin{tabular*}{\columnwidth}{@{\extracolsep{\fill}}lcc@{}}
\toprule
\textbf{Variant} &
\textbf{Overall (\%)} &
\textbf{$\Delta$ (pp)} \\
\midrule
\textbf{XferAgent (Full)} &
\textbf{52.16} & -- \\
\midrule
Local Experience Only &
46.84 & -5.32 \\
Shared Experience Only &
50.83 & -1.33 \\
\addlinespace[2pt]
w/o Outcome-Based Ranking &
48.17 & -3.99 \\
w/o Connectivity Constraint &
46.84 & -5.32 \\
\bottomrule
\end{tabular*}
\end{table}

\subsection{Sensitivity Analysis}
\label{sec:sensitivity}


We examine four key parameters of XferAgent's experience reconciliation, routing, and integration on 45 held-out GAIA validation tasks using GPT-5.6 Luna, with experience graphs constructed from the remaining 120 tasks. The default settings are $B=24$ (subgraph budget), $K=3$ (final experience limit), $\lambda_u=0.40$ (ranking weight), and $\delta_{\mathrm{rec}}=0.03$ (reconciliation margin). Each parameter is varied independently, with the others fixed. Reconciliation variants replay the recorded graph construction trajectory, while other configurations reuse the same constructed memories. Further protocol details are provided in Appendix~\ref{app:efficiency}.


\begin{table}[t]
\centering
\caption{Parameter sensitivity of XferAgent on 45 held-out GAIA validation tasks using GPT-5.6 Luna. One parameter is varied at a time, with others fixed at their default values. $\Delta$ denotes the accuracy change relative to the default in percentage points. 
}
\label{tab:sensitivity_validation}
\small
\setlength{\tabcolsep}{4pt}
\renewcommand{\arraystretch}{0.9}
\begin{tabular*}{\columnwidth}{@{\extracolsep{\fill}}cccc@{}}
\toprule
\textbf{Parameter} &
\textbf{Value} &
\textbf{Overall (\%)} &
\textbf{$\Delta$ (pp)} \\
\midrule
\multirow{3}{*}{$B$}
 & 8 & 51.11 & -6.67 \\
 & 24 (default) & 57.78 & -- \\
 & 40 & 46.67 & -11.11 \\
\midrule
\multirow{3}{*}{$K$}
 & 1 & 51.11 & -6.67 \\
 & 3 (default) & 57.78 & -- \\
 & 5 & \textbf{62.22} & +4.44 \\
\midrule
\multirow{3}{*}{$\lambda_u$}
 & 0 & 53.33 & -4.44 \\
 & 0.40 (default) & 57.78 & -- \\
 & 1 & 55.56 & -2.22 \\
\midrule
\multirow{3}{*}{$\delta_{\mathrm{rec}}$}
 & 0 & 51.11 & -6.67 \\
 & 0.03 (default) & 57.78 & -- \\
 & 0.08 & 48.89 & -8.89 \\
\bottomrule
\end{tabular*}
\vspace{-1em}
\end{table}




Increasing the final experience limit to $K=5$ yields the highest observed accuracy of 62.22\% (Table~\ref{tab:sensitivity_validation}), while the default ranking weight $\lambda_u=0.40$ outperforms both tested extremes. Changes to $B$ and $\delta_{\mathrm{rec}}$ do not alter the logged routing decisions or final experience selections, so their accuracy differences do not establish retrieval-related improvements. These comparisons characterize the observed parameter sensitivity rather than identifying optimal settings.


\FloatBarrier

\section{Conclusion}

We introduced \textbf{XferAgent}, a framework for Cross-Agent Experience Transfer that enables heterogeneous LLM agents to collaboratively reuse experiences beyond their individual memories. Experiments across diverse tasks demonstrate improvements over representative memory-based approaches, while repeated-task evaluations show increasing cumulative success through experience evolution. XferAgent provides a foundation for experience-level collaboration toward self-improving agent ecosystems.

\clearpage

\section*{Limitations}
For cross-agent experience sharing, XferAgent removes task-specific and execution-sensitive information from contributed experiences while keeping raw tasks and task embeddings on the agent side. These measures limit the information shared but do not provide formal privacy guarantees. XferAgent currently focuses on text-based reusable experiences, and future work may extend the framework to more complex multimodal agent settings. Separating source and evaluation tasks does not rule out semantic overlap or exposure during model pretraining.




\bibliography{custom}

\appendix

\section{Algorithm and Implementation Details}
\label{app:experience_algorithm}

Algorithm~\ref{alg:fedagentke_expgraph} describes XferAgent's routing,
integration, execution, and experience updates. \textsc{Reconcile} maps
new contributions into the Shared Experience Graph.
\textsc{RouteConnected} applies PPR and grows a connected Routed
Experience Subgraph under the node budget.
\textsc{IntegrateAndSelect} expands the local and routed graphs,
constructs a temporary fused graph, and selects experiences for the prompt.

Here, $i$ identifies the target agent for outcome and contributor
statistics; $f_i$ identifies its execution framework. $D_i$ maps a completed
execution to an experience node, and $P_i$ produces its shareable
representation. $\mathcal I_i$ contains available node identifiers and
embeddings; $\mathcal A_i$ contains anchors and their task similarities.
$\mathcal B_i$ is the Routed Experience Subgraph, and $\mathcal R_{i,t}$
contains the experiences selected for the task prompt.

\begin{algorithm}[!ht]
\caption{XferAgent with Task-Conditioned Experience Routing}
\label{alg:fedagentke_expgraph}
\small
\begin{algorithmic}[1]
\Require Local Experience Graphs $\{\mathcal G_i^L\}$, Shared Experience Graph $\mathcal G^S$,
node budget $B$, frozen configuration $\Gamma$, evaluation mode
\For{target task $x_t$ assigned to agent $i$ in commit order}
  \State $\mathcal I_i\gets$ available node identifiers and embeddings
  from $\mathcal G^S$ under the authorization policy for $f_i$
  \State $\mathcal A_i\gets\textsc{AgentAnchors}(x_t,\mathcal I_i;\Gamma)$
  \State $\mathcal B_i\gets
  \textsc{RouteConnected}(\mathcal G^S,\mathcal A_i,i,f_i,B;\Gamma)$
  \If{the Shared Experience Graph state changed during routing}
    \State Repeat lines~2--4 using the updated Shared Experience Graph
  \EndIf
  \State $\mathcal R_{i,t}\gets
  \textsc{IntegrateAndSelect}(x_t,\mathcal G_i^L,\mathcal B_i,i;\Gamma)$
  \State Record the Routed Experience Subgraph and exact local and shared
  experience nodes represented in $\mathcal R_{i,t}$
  \State $y_t\gets\operatorname{Agent}_{f_i}(x_t;\mathcal R_{i,t})$
  \State $r_t\gets Q(x_t,y_t)$
  \If{experience updates are enabled}
  \State Update outcome statistics only for experiences in the prompt:
  local observations and agent-$i$ records in the Shared Experience Graph
  \State $\widetilde v_{i,t}\gets D_i(x_t,y_t,r_t)$
  \State $\mathcal G_i^L\gets
  \textsc{InsertAndRebuild}(\mathcal G_i^L,\widetilde v_{i,t};\Gamma)$
  \State $\overline v_{i,t}\gets P_i(\widetilde v_{i,t})$
  \State $(\mathcal G^S,v^S)\gets
  \textsc{Reconcile}(\mathcal G^S,\overline v_{i,t},i;\Gamma)$
  \State Add a \textsc{DERIVED-FROM} relation from $v^S$ to each shared
  experience node included in the task prompt
  \EndIf
\EndFor
\State \Return $\{\mathcal G_i^L\}$ and $\mathcal G^S$
\end{algorithmic}
\end{algorithm}

\paragraph{Evaluation and updates.}
Frozen-memory evaluation runs routing, integration, execution, and grading
without updating statistics, constructing experiences, or writing back to
the graphs. When updates are enabled, only experiences included in the
prompt receive outcome-statistics updates. A failed task may lower their
scores and produce a new \textsc{FAILURE} experience. A task whose prompt contains no
experiences leaves prior statistics unchanged but may still
produce a new experience after grading.

\paragraph{Routing interface.}
Before routing, the agent receives only available node identifiers and
embeddings; experience text, provenance, outcome statistics, contributor
identities, and graph structure remain in the Shared Experience Graph
until the routed subgraph is delivered. \textsc{AgentAnchors} computes
the task embedding locally. Positive anchor similarities are normalized
to form the PPR restart distribution, and isolated nodes receive self-loops.

\paragraph{Local--Shared integration.}
Shared candidates initialize the pool. Local candidates are processed in
descending source-graph PPR order, matching first by a recorded shared
identifier when its candidate is present, otherwise by similarity among all
same-type candidates already in the pool. This includes earlier local
candidates. Each match preserves any shared identifier but replaces the
text, embedding, and local statistics with those of the incoming local
candidate. The last local match therefore supplies the representative;
local statistics are not pooled. Task similarity retains the maximum
pre-merge value rather than being recomputed for the retained text.

Edges are remapped to the fused nodes, discarding self-loops and taking
the maximum weight for duplicate edges. Additional semantic edges connect
only local-only and shared-only candidates, retaining the most similar
neighbors subject to the similarity
threshold and neighbor limit. PPR on this temporary graph uses the
highest-similarity candidates meeting the task-matching threshold as
anchors, up to the anchor limit. Positive similarities supply the
normalized restart weights; without them, retrieval returns no experiences.
Candidates with positive PPR scores are retained in descending order up
to the candidate limit before final ranking. Stable identifiers resolve
sorting and matching ties throughout integration.

\paragraph{Scoring and normalization.}
All outcome scores use pre-task statistics. Candidates with both local
and shared records combine local outcome sums and usage counts with the
cross-agent outcome score as a prior (Eq.~\ref{eq:local_global_utility}). Local-only candidates use
$u_0$ as the prior; shared-only candidates use the cross-agent score directly.

The exploration bonus is $c\sqrt{\log(1+T_{\mathrm{pool}})/\max(1,n_v)}$,
where $c$ is fixed, $n_v$ is the node's usage count, and
$T_{\mathrm{pool}}$ is the total usage count in the relevant pool.
Routing uses available shared nodes and excludes the target agent's own
uses. The bonus is added to $s^{\mathrm{outcome}}_{-i,v}$ and capped at one
to obtain $s^{\mathrm{exp-outcome}}_{-i,v}$. Contributor support
$s^{\mathrm{support}}_{-i,v}$ counts distinct other contributing agents,
measuring provenance breadth rather than successful uses.

Final ranking uses only candidates retained by fused-graph PPR. Candidates
with both local and shared records combine local usage counts with other
agents' shared counts;
local-only and shared-only candidates use their available counts. The
bonus is added to $s^{\mathrm{fused}}_{i,v}$ and capped at one to obtain
$s^{\mathrm{exp-fused}}_{i,v}$. Each score term is min-max normalized within
its routing or final-ranking pool. If the range is numerically zero,
the original values are retained after clipping to $[0,1]$.

Benchmark settings are specified with their adapters: Table~\ref{tab:alfworld_graph_configuration}
reports the ALFWorld graph configuration, Appendix~\ref{app:common_benchmark_protocol}
gives MBPP memory and inference limits, and Appendix~\ref{app:efficiency}
defines the GAIA sensitivity defaults.

\section{GAIA Evaluation Details}
\label{app:gaia_test_backbones}

We report GAIA 2023 Test results for two reasoning backbones in
Table~\ref{tab:main_overall}.
The Test set contains 301 tasks: 93 at Level~1, 159 at
Level~2, and 49 at Level~3.  Correct-answer counts come from the official
hidden-label evaluation.  Memories remain frozen during GAIA Test evaluation.

\paragraph{Memory construction and execution.}
The validation source pool contains 165 tasks: 53 at Level~1, 86 at
Level~2, and 26 at Level~3. The three-client GAIA adapters assign each
level to a logical client with its own local memory. Shared retrieval
accesses experiences published by other clients. Test tasks do not supply
construction feedback, and no model weights are fine-tuned.

Construction is method-specific. Qwen ReasoningBank extracts strategy
memories from cached AgentKB validation trajectories generated with the
same backbone. Luna ReasoningBank and Mem0 collect fresh validation
trajectories. Memory size varies with the construction method and
completed source records.

Qwen3-Next-80B-A3B-Instruct is served through Vertex MaaS;
GPT-5.6 Luna uses the OpenAI API. The GAIA execution interface uses
SmolAgents with search, browser, document, and Python tools.
Overall accuracy uses all 301 tasks; level-wise accuracy uses 93, 159,
and 49 tasks for Levels~1--3, respectively. Failed tasks remain in their
corresponding denominators.
Blank answers count as incorrect; nonempty submitted answers are scored
by the official evaluator, including answers retained after execution errors.

Table~\ref{tab:gaia_test_by_level} provides the difficulty-level breakdown
of the GAIA Test results in Table~\ref{tab:main_overall}.

\begin{table}[t]
\centering
\small
\caption{GAIA Test accuracy (\%) by difficulty. Bold marks the best score in each column within each model block, including ties.}
\label{tab:gaia_test_by_level}
\setlength{\tabcolsep}{4pt}
\renewcommand{\arraystretch}{1.08}
\begin{tabular*}{\columnwidth}{@{\extracolsep{\fill}}lrrrr@{}}
\toprule
\multicolumn{5}{c}{\textbf{Qwen3-Next-80B-A3B-Instruct}} \\
\midrule
\textbf{Method} & \textbf{Overall} & \textbf{L1} & \textbf{L2} & \textbf{L3} \\
\midrule
No Memory     & 27.57 & 41.94 & \textbf{26.42} & 4.08 \\
AgentKB       & 28.90 & \textbf{49.46} & 22.64 & 10.20 \\
ReasoningBank & 26.91 & 43.01 & 23.27 & 8.16 \\
ExpGraph      & 27.24 & 44.09 & 24.53 & 4.08 \\
Mem0          & 25.91 & 41.94 & 23.27 & 4.08 \\
LightMem      & 21.59 & 35.48 & 16.98 & 10.20 \\
XferAgent     & \textbf{29.57} & 47.31 & 24.53 & \textbf{12.24} \\
\midrule
\multicolumn{5}{c}{\textbf{GPT-5.6 Luna}} \\
\midrule
\textbf{Method} & \textbf{Overall} & \textbf{L1} & \textbf{L2} & \textbf{L3} \\
\midrule
No Memory     & 48.17 & 66.67 & 45.91 & 20.41 \\
AgentKB       & 46.51 & 60.22 & 46.54 & 20.41 \\
ReasoningBank & 46.18 & 59.14 & 45.91 & 22.45 \\
ExpGraph      & 47.51 & 68.82 & 41.51 & \textbf{26.53} \\
Mem0          & 44.85 & 65.59 & 40.88 & 18.37 \\
LightMem      & 41.86 & 52.69 & 42.77 & 18.37 \\
XferAgent     & \textbf{52.16} & \textbf{72.04} & \textbf{48.43} & \textbf{26.53} \\
\bottomrule
\end{tabular*}
\end{table}

\paragraph{Progressive evaluation.}
Section~\ref{sec:cross_round} evaluates cumulative recovery over three
passes on the same 165 validation tasks (53/86/26 at Levels~1--3).
The first pass starts with empty experience memories. Each subsequent
pass re-evaluates all 165 tasks, including those solved previously.
Experience updates follow task grading, and memories are carried forward
between passes. Later attempts may therefore retrieve experiences derived
from earlier attempts on the same task. A task is counted as correct if it has been solved at least once by the
end of the current pass; each task contributes at most once, and the
denominator remains 165. This protocol combines repeated attempts with
experience updates. No Memory provides an independent reference rather
than a matched three-pass control.

\section{Additional Evaluation Protocols}
\label{app:supplementary_protocols}

Table~\ref{tab:additional_protocol_overview} summarizes the ALFWorld,
MBPP, and cross-framework GAIA protocols. Memory methods use
benchmark-specific adapters with agent-owned memories and bounded
cross-agent retrieval. Memories are constructed from collected
trajectories without fine-tuning model weights.

Within each benchmark, models, evaluation tasks, and retrieval limits are
matched across methods. ALFWorld uses a fixed cohort across adaptations.
ALFWorld and MBPP rates include failed and unparsed cases and are rounded
to one decimal place.

\begin{table*}[t]
\centering
\small
\caption{Settings for the additional evaluations. Memory content and outcome statistics are frozen during evaluation. Memory allowances specify the token limits for retrieved context.}
\label{tab:additional_protocol_overview}
\setlength{\tabcolsep}{4pt}
\renewcommand{\arraystretch}{1.12}
\begin{tabular}{@{}p{0.18\textwidth}p{0.25\textwidth}p{0.25\textwidth}p{0.25\textwidth}@{}}
\toprule
\textbf{Setting} & \textbf{ALFWorld} & \textbf{MBPP} & \textbf{Cross-framework GAIA} \\
\midrule
Backbone & Qwen3.8-27B (Tinker) & Llama-3.1-8B-Instruct & GPT-5.6 Luna \\
Construction data & 90 training trajectories & 265 cached training trajectories & 20 questions per agent from a 105-question source pool \\
Evaluation cohort & 274 tasks: 140 Seen, 134 Unseen & 60 development tasks & 60 tasks: 18 / 29 / 13 at Levels 1--3 \\
Agent arrangement & Three logical agents & Three logical agents & Nested 2 / 4 / 8 contributors; one fixed receiver \\
Shared budget $B$ / final limit $K$ & 8 / 10 & 8 / 10 & 24 / 3 \\
Memory allowance & 2,400 tokens & 2,400 tokens (character estimate) & 12,000 tokens \\
\bottomrule
\end{tabular}
\end{table*}

\subsection{MBPP: Coding Development Subset}
\label{app:common_benchmark_protocol}

The evaluation cohort in Table~\ref{tab:additional_protocol_overview}
contains 60 tasks from a mixed MBPP/MBPPPlus development pool of 100 tasks.
Cached construction trajectories have a non-IID three-agent assignment.
The memory-construction and evaluation sets are disjoint at the task level.

\paragraph{Inference and memory settings.}
MBPP uses greedy decoding, per-call limits of 4,096 input and 2,048 output
tokens, and a single beginning-of-sequence token.
Dense retrieval uses the frozen Contriever encoder with a 256-token
encoding limit. Agent ownership and evaluation IDs are fixed within each
comparison; inference and memory limits appear in
Table~\ref{tab:additional_protocol_overview}.

\paragraph{Experience-sharing baseline adapters.}
For MBPP, the memory-based methods use agent-owned memories and bounded
cross-agent retrieval; No Memory evaluates without retrieval. AgentKB ranks up to
10 local and eight remote candidates by dense similarity and includes at
most 10 entries in the prompt. Table~\ref{tab:mbpp_adapter_scope} summarizes
the method-specific adaptations. Agents are logical partitions with
in-process routing; network efficiency is outside the evaluation scope.

\paragraph{Evaluation.}
Each task produces one final program, extracted with the frozen parser
and evaluated using the stored \texttt{check(candidate)} harness with a
30-second timeout. Pass rate is computed using this stored harness
rather than the full-benchmark EvalPlus suite.

\begin{table}[t]
\centering
\small
\caption{Memory adapters used for MBPP.}
\label{tab:mbpp_adapter_scope}
\setlength{\tabcolsep}{4pt}
\renewcommand{\arraystretch}{1.08}
\begin{tabular}{@{}lp{0.65\columnwidth}@{}}
\toprule
\textbf{Method} & \textbf{Benchmark-specific adaptation} \\
\midrule
AgentKB & Flat workflow entries; bounded local and cross-client similarity retrieval. \\
ReasoningBank & Dense retrieval over cached training experiences. \\
ExpGraph & Graph retrieval over cached training experiences, frozen during evaluation. \\
Mem0 & Source-preserving procedural-memory updates in a persistent store. \\
LightMem & Short-term grouping and offline consolidation with source text retained. \\
\bottomrule
\end{tabular}
\end{table}

\subsection{ALFWorld: Embodied Task Completion}

ALFWorld evaluates language-guided household tasks through a textual
environment. Success rate is pooled over the Seen and Unseen cohort in
Table~\ref{tab:additional_protocol_overview}, using the final environment
state. Thinking is disabled and temperature is zero, with per-call limits
of 8,192 input and 256 output tokens. The action interface exposes legal actions. Generated actions are passed
to the environment unchanged, including invalid actions. Retrieval uses
a training-derived TF-IDF adaptation for the reported memory and graph methods.

The 90 source trajectories were generated with Qwen3.5-9B and reprocessed
by the 27B model, yielding 74 AgentKB entries, 90 ReasoningBank entries,
and 80 XferAgent entries.

Table~\ref{tab:alfworld_graph_configuration} lists the key settings.
The ALFWorld adapter additionally checks Jaccard overlap between task-facet
sets derived from task keywords and capability tags before merging.
Nonempty tool sets must overlap, as must nonempty failure-condition sets
for \textsc{failure} experiences. During connected expansion, anchor rewards use task
similarities, edge rewards use edge weights, and connector costs sum
$1-w$ along the path, with edge weights clipped to $[0,1]$.
The disclosure-risk term is inactive: all frozen shared nodes have zero
risk scores.

No Memory uses no retrieved experience. The AgentKB adapter uses a flat
knowledge base partitioned across three logical agents.
Each episode retrieves once at the start, using at most 10 private
candidates from the target agent and eight cross-agent candidates, then at most
10 final entries within 2,400 tokens. It uses neither graph structure
nor outcome-based ranking, and its KB is frozen throughout testing.

The other adapters also use the 90 raw training trajectories.
ReasoningBank uses local and cross-agent TF-IDF similarity over the
re-extracted memories. ExpGraph uses diffusion within agent-owned graphs. Mem0 applies
ADD/UPDATE/DELETE operations to agent-owned memories; LightMem extracts
short-term lessons and consolidates them within each agent.

\begin{table}[t]
\centering
\small
\caption{Key settings of the ALFWorld XferAgent adapter.}
\label{tab:alfworld_graph_configuration}
\setlength{\tabcolsep}{4pt}
\renewcommand{\arraystretch}{1.08}
\begin{tabular*}{\columnwidth}{@{\extracolsep{\fill}}lr@{}}
\toprule
\textbf{Setting} & \textbf{Value} \\
\midrule
Shared node budget / final limit & 8 / 10 \\
Memory allowance (tokens) & 2,400 \\
Experience-text limit (characters) & 5,000 \\
Reconciliation threshold / margin & 0.92 / 0 \\
Structural-overlap threshold & 0.50 \\
Anchor limit / task-match threshold & 10 / 0.08 \\
Expansion / fused-candidate limits & 20 / 20 \\
Fusion-merge threshold & 0.92 \\
PPR restart probability & 0.35 \\
Outcome prior $u_0$ / strength $\kappa$ & 0.50 / 2 \\
Exploration weight $c$ & 0.25 \\
Final outcome weight & 0.40 \\
Routing: relevance / outcome / support & 0.55 / 0.25 / 0.15 \\
Edge / anchor / connector weights & 0.10 / 0.25 / 0.10 \\
\bottomrule
\end{tabular*}
\end{table}

\subsection{Cross-Framework Transfer Protocol}
\label{app:cross_framework_protocol}
\paragraph{Data and agent groups.}
The 165 GAIA validation questions are split into the source and evaluation
cohorts in Table~\ref{tab:additional_protocol_overview}.
Table~\ref{tab:cross_framework_groups} gives the contributor groups.
Larger groups retain the smaller groups' agents and constructed memories.
Each agent samples its source questions uniformly without replacement;
samples may overlap across agents.

\begin{table}[!ht]
\centering
\small
\caption{Nested contributor groups for cross-framework transfer. Each contributor builds memory from 20 source questions. Totals count agent--question instances; questions may overlap across agents.}
\label{tab:cross_framework_groups}
\setlength{\tabcolsep}{4pt}
\begin{tabular*}{\columnwidth}{@{\extracolsep{\fill}}rrrr@{}}
\toprule
\textbf{Agents} & \textbf{SmolAgents} & \textbf{OWL} & \textbf{Total instances} \\
\midrule
2 & 1 & 1 & 40 \\
4 & 2 & 2 & 80 \\
8 & 4 & 4 & 160 \\
\bottomrule
\end{tabular*}
\end{table}

\paragraph{Construction and evaluation.}
Each agent constructs local memory from its selected source tasks;
contributions are imported in a fixed agent order to construct the Shared
Experience Graph. The receiver is the first SmolAgents contributor, using
the same local memory in every condition.
It answers all 60 evaluation questions, producing one final answer per
question without multi-agent voting.

Model calls allow up to three API attempts with a 120-second request
timeout and no additional SDK retries; the task time limit is 2,700 seconds.
The figure's Average is the unweighted mean of the three level accuracies,
not pooled accuracy over the 60 questions. Agent count and aggregate
learning exposure increase together.

\section{Sensitivity Protocol}
\label{app:efficiency}

The sensitivity study in Section~\ref{sec:sensitivity} uses GPT-5.6 Luna
with 120 GAIA validation tasks for memory construction and 45 disjoint
held-out tasks for evaluation. Construction contains 38, 63, and 19
questions at Levels~1--3; evaluation contains 15, 23, and 7, respectively.
One logical agent is assigned to each level.

Each variant changes one factor from the default: $B=24$, $K=3$,
$\lambda_u=0.40$, and reconciliation margin $0.03$. $K$ limits the
experiences included in the prompt; $\lambda_u$ weights the outcome term
in final ranking, not shared graph routing.

Table~\ref{tab:sensitivity_validation} reports nine configurations:
one default and eight variants. The default rows repeat the same
45-task result, not separate evaluations.

Margin variants reconstruct the shared graph by replaying the original
construction and feedback events without new solver calls. Feedback is
assigned to corresponding reconstructed nodes available at each replay
step, counting each mapped node at most once per historical exposure.
Local graphs and their identifiers remain unchanged; integration follows
Appendix~\ref{app:experience_algorithm}. Other variants reuse the original
graphs. Memories and outcome statistics are frozen during evaluation;
all remaining settings are held fixed.

Each task contributes one outcome per configuration; failures count as
incorrect in the 45-task denominator.

\end{document}